\documentclass[aps,prl,reprint,groupedaddress,nofootinbib]{revtex4-1}
\usepackage{hyperref}
\usepackage{amsmath}
 \usepackage{multirow}
\usepackage{array}
\newcolumntype{L}[1]{>{\raggedright\let\newline\\\arraybacksslash\hspace{0pt}}m{#1}}
\newcolumntype{C}[1]{>{\centering\let\newline\\\arraybackslash\hspace{0pt}}m{#1}}
\newcolumntype{R}[1]{>{\raggedleft\let\newline\\\arraybackslash\hspace{0pt}}m{#1}}
\usepackage{float}
\usepackage{graphicx}
\usepackage{epsfig}
\usepackage{psfrag}
\usepackage{color}
\usepackage{slashed}

\usepackage{amsfonts}
\usepackage{amssymb}
\usepackage{tikz}
\usepackage{tikz}
\usetikzlibrary{positioning,arrows}
\usetikzlibrary{decorations.pathmorphing}
\usetikzlibrary{decorations.markings}

\newcommand*{\be}{\begin{equation}}
\newcommand*{\ee}{\end{equation}}
\newcommand*{\bea}{\begin{eqnarray}}
\newcommand*{\eea}{\end{eqnarray}}

\newcommand{\comment}[1]{}

\newcommand{\cref}[1]{Chapter~\ref{c.#1}}

\def\beq{\begin{equation}}
\def\eeq{\end{equation}}
\def\bea{\begin{eqnarray}}
\def\eea{\end{eqnarray}}
\def\ba{\begin{array}}
\def\ea{\end{array}}
\def\bi{\begin{itemize}}
\def\ei{\end{itemize}}
\def\be{\begin{enumerate}}
\def\ee{\end{enumerate}}
\def\bc{\begin{center}}
\def\ec{\end{center}}
\def\bt{\begin{table}}
\def\et{\end{table}}
\def\btb{\begin{tabular}}
\def\etb{\end{tabular}}

\def\lsim{\raise0.3ex\hbox{$\;<$\kern-0.75em\raise-1.1ex\hbox{$\sim\;$}}}
\def\gsim{\raise0.3ex\hbox{$\;>$\kern-0.75em\raise-1.1ex\hbox{$\sim\;$}}}
	
\usepackage{cancel}
 \usepackage[normalem]{ulem}
\usepackage{color}
\def\comment#1{\textcolor{blue}{\large(\it{#1})}}

\def\lapp{\mathrel{\rlap{\raise.5ex\hbox{$<$}}
                    {\lower.5ex\hbox{$\sim$}}}}
\def\gapp{\mathrel{\rlap{\raise.5ex\hbox{$>$}}
                    {\lower.5ex\hbox{$\sim$}}}}

\begin{document}

\title{Tera-Zooming in on light (composite) axion-like particles}

\author{G. Cacciapaglia$^{1,2}$}
\email{g.cacciapaglia@ipnl.in2p3.fr}
\author{A. Deandrea$^{1,2}$}
\email{deandrea@ipnl.in2p3.fr}
\author{A.M.  Iyer$^{1,2,5}$}
\email{iyerabhishek@physics.iitd.ac.in ,a.iyer@ipnl.in2p3.fr}
\author{K. Sridhar$^{3,4}$}
\email{sridhar@theory.tifr.res.in}

\affiliation{
$^1$University of Lyon, Universit\'e Claude Bernard Lyon 1,
F-69001 Lyon, France\\
$^2$Institut de Physique des 2 Infinis de Lyon (IP2I), UMR5822, CNRS/IN2P3,  F-69622 Villeurbanne Cedex, France\\
$^3$ Department of Theoretical Physics, Tata Institute of Fundamental Research, Mumbai-400005, India\\
$^4$Krea University, Sri City, Andhra Pradesh-517646, India \\
$^5$ Department of Physics, Indian Institute of Technology Delhi, New Delhi-110016, India}


\begin{abstract}
The Tera-Z phase of future $e^+ e^-$ colliders, FCC-ee and CepC, is a goldmine for exploring $Z$ portal physics. We focus on axion-like particles (ALPs) that can be produced via $Z$ decays with a monochromatic photon. As a template model, we consider composite Higgs models with a light pseudo-scalar that couples through the Wess-Zumino-Witten term to the electroweak gauge bosons. For both photophilic and photophobic cases, we show that the Tera-Z can probe composite scales up to $100$s of TeV, well beyond the capability of the LHC and current precision physics.
Our results also apply to generic ALPs and, in particular, severely constrain models that explain the muon $g-2$ anomaly.

\end{abstract}

\maketitle

\noindent 

New physics extensions of the Standard Model (SM) often contain additional scalar bosons. Their presence can be linked to symmetries, as is the case for axions~\cite{PhysRevLett.38.1440,PhysRevD.16.1791} and for the second Higgs doublet in supersymmetry~\cite{Haber:1984rc}, or to the composite nature of the SM extension~\cite{Kaplan:1983sm,Dugan:1984hq,Georgi:1986im}. Extended scalar sectors are also popular as they offer interesting model building avenues: multiple doublets~\cite{Lee:1973iz,Branco_2012} enrich the Higgs sector of the SM, while triplets emerge in neutrino type-II see-saw models~\cite{Mohapatra:1980yp} and the custodial Georgi-Machacek model~\cite{Georgi:1985nv,Chanowitz:1985ug}. Singlets can provide Dark Matter candidates \cite{Barger:2007im,Barger:2008jx} or help achieving an electroweak first order phase transition \cite{Profumo:2007wc,Espinosa:2011ax}, as required by baryogenesis and leptogenesis \cite{Davidson:2008bu}. 
Pseudo-scalars are of special interest, as they exhibit different properties from those of the Higgs boson. While these states map to several Beyond the Standard Model (BSM) scenarios, they can be broadly classified on the basis of the nature of their interactions. 
In this letter, we focus on Axion-Like Particles (ALPs)~\cite{Jaeckel:2010ni,Arias:2012az}, which, being gauge singlets, couple to the SM gauge bosons and fermions via dimension-5 operators. They have had implications for and have been studied in several areas of particle physics: flavour \cite{Marciano:2016yhf,Bauer:2017nlg,Bauer:2019gfk,Calibbi:2016hwq,Calibbi:2020jvd,Chakraborty:2021wda,Cacciapaglia:2020kbf,Izaguirre:2016dfi,Cornella:2019uxs}, colliders \cite{Bauer:2017ris,Cacciapaglia:2019bqz,Cacciapaglia:2017iws,Belyaev:2015hgo,Ebadi:2019gij,Gavela:2019cmq,Brivio:2017ije}, as a dark matter candidate \cite{Cacciapaglia:2020kbf,Cai:2019cow,Alonso-Alvarez:2019ssa,Daido:2017wwb,Arias:2012az,Brivio:2015kia,Arbey:2015exa,Belyaev:2016ftv}. For a comprehensive  insight into these models, see for example \cite{Cacciapaglia:2020kgq,Brivio:2017ije,Bauer:2020jbp}.

In particular, motivated by composite Higgs models, we will study a special case where only Wess-Zumino-Witten (WZW) couplings \cite{Wess:1971yu,Witten:1983tw} to the electroweak gauge bosons are present at the leading order:
\begin{equation}
\mathcal{L}_{WZW}=a \left(g^2\frac{C_W}{\Lambda} W_{\mu\nu}\tilde W^{\mu\nu}+{g'}^{2}\frac{C_B}{\Lambda} B_{\mu\nu}\tilde B^{\mu\nu} \right)\,,
\label{eq:WZW}
\end{equation}
where $W,\ B$ are the electroweak bosons of the $SU(2)_L \times U(1)_Y$ gauge group respectively and $a$ is the pseudo-scalar field. Here, $\Lambda$ is the mass scale in the effective theory where the couplings are generated, while the coefficients $C_{W,B}$ are determined by the ultraviolet (UV) completion. 
While these couplings could be embedded in several UV frameworks, two scenarios of interest are ALPs \cite{Jaeckel:2010ni,Arias:2012az} and models of composite Higgs \cite{Dugan:1984hq,Galloway:2010bp,Arbey:2015exa,Belyaev:2015hgo}. In either case, the pseudo-scalar $a$ emerges as a pseudo Nambu-Goldstone Boson (pNGB) of a spontaneously broken global symmetry and, therefore, can be much lighter that the scale $\Lambda$. Due to the absence of couplings to gluons and fermions, direct and indirect bounds are relatively weak, especially in the mass range between $1$~GeV and the $Z$ mass (e.g., see~\cite{Bauer:2017nlg,Bauer:2017ris,Craig:2018kne}). Couplings to fermions are, in fact, generated at loop level~\cite{Bauer:2017ris}, and their effect is fully taken into account in this work.

In this letter we point out that future electron-positron colliders running at the $Z$ pole, like FCC-ee~\cite{Abada:2019zxq} and CepC~\cite{CEPCStudyGroup:2018ghi}, are discovery machines for these light $a$'s thanks to the leading-order coupling $Z a \gamma$ contained in Eq.~\eqref{eq:WZW}. With a projected number of $Z$ bosons produced in the few times $10^{12}$ ballpark, the so-called Tera-Z run has the capability of probing the composite nature of the Higgs boson up to very large scales. In this work, we will focus on two distinct scenarios:
\begin{itemize}
    \item[a)] {\bf Photophobic}, corresponding to $C_B = - C_W$ in Eq.~\eqref{eq:WZW}. In this case, the coupling to photons ($C_{\gamma \gamma} = C_W + C_B$) vanishes. Composite models based on $SU(4)/Sp(4)$~\cite{Kaplan:1983sm,Galloway:2010bp,Arbey:2015exa} and $SU(4)^2/SU(4)$~\cite{Ma:2015gra} fall under this category. Current bounds have been collected in~\cite{Craig:2018kne}.
    \item[b)] {\bf Photophilic}, corresponding to $C_B=C_W$. In this case, the dominant coupling for masses below the $Z$ is to two photons. Composite models based on $SU(5)/SO(5)$~\cite{Dugan:1984hq,Ferretti:2014qta,Agugliaro:2018vsu} fall in this category. The current bounds mainly relying on the photon coupling are summarised in~\cite{Bauer:2017nlg}. We should also note that loop induced couplings to the Higgs in composite models are suppressed once compared to generic ALP scenarios~\cite{Cacciapaglia:2017iws}.  
\end{itemize}
A remarkable class of composite ALPs, which falls in the Photophilic class, is composed of $U(1)$ pNGBs in models with top partial compositeness~\cite{Belyaev:2015hgo,Belyaev:2016ftv,Cacciapaglia:2019bqz}: while they feature leading order couplings to gluons, the LHC bounds remain relatively weak~\cite{Cacciapaglia:2019bqz}. As already mentioned, the results we present in this work are not special to composite models and can be applied straightforwardly to a generic ALP scenario.
We use the composite Higgs framework as a template to visualize the results. As such, the coupling $C_W$ can be expressed as
\begin{equation} \label{eq:CW}
    \frac{C_W}{\Lambda} = \frac{d_\psi}{64 \sqrt{2} \pi^2 f}\,,
\end{equation}
where $f$ is the decay constant of the composite Higgs (it is implicitly assumed that $f$ is larger than the Higgs vacuum expectation value $v$) and $d_\psi$ counts the internal degrees of freedom of the fermions confined in the composite Higgs. In the numerical study, we will fix $d_\psi = 4$ following the most minimal cases~\cite{Ferretti:2013kya}. The results will also be shown in terms of the so-called fine-tuning parameter $v/f$.

\begin{figure}[htb!]
	\centering	
	\begin{tabular}{c}
		\includegraphics[width=7.5cm]{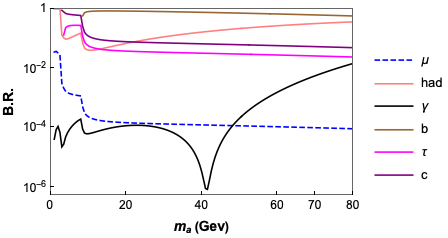}
	\end{tabular}
	\caption{Branching ratios (B.R.) for the pseudo-scalar $a$ in the Photophobic case, as a function of its mass.}
	\protect\label{fig:BF}
\end{figure}

\begin{figure*}[htb!]
	\centering	
	\begin{center}
		\begin{tabular}{c}
			\includegraphics[width=18cm]{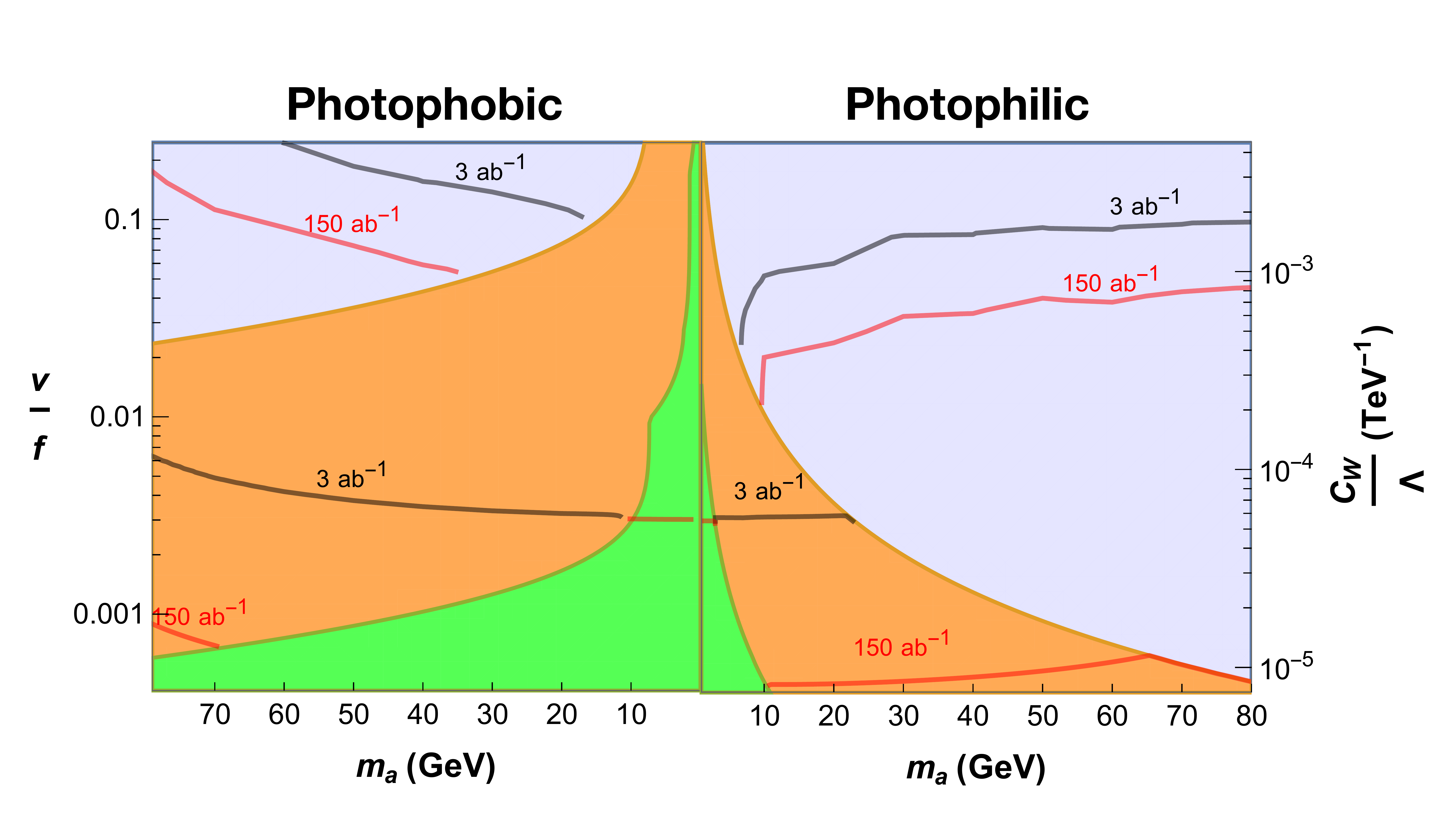}
		\end{tabular}
	\end{center}
	\caption{Butterfly plot showing the parameter space for the photophilic (right wing) and photophobic (left wing) models. The shaded regions correspond to prompt decays (blue), long-lived (orange) and missing energy signature (green). The $2\sigma$ reach of the Tera-Z run is indicated for integrated Luminosities of $3$ (black) and $150$~ab$^{-1}$ (red). The conversion between the ALP coupling $C_W/\Lambda$ and the compositeness scale $f$ is given in Eq.~\eqref{eq:CW} with $d_\psi = 4$.}
	\protect\label{fig:results}
\end{figure*}

We use the results from \cite{Bauer:2017ris}, including loop-induced couplings to fermions and photons, to calculate the partial decay widths of the pseudo-scalar $a$. As all the couplings are proportional to the same coefficient $C_W/\Lambda$, the branching ratios only depend on the $a$ mass, as illustrated in Fig.\ref{fig:BF} for the Photophobic case. The dominant decay is always into the heaviest accessible quark, while sub-leading decays into photons grow towards the $Z$ mass threshold. In the Photophilic Case, the dominant decay is always into two photons, independent of the mass.
The only physical quantities that depend on the WZW coupling are the branching ratio of $Z \to a \gamma$ and the lifetime of $a$. The former fixes the number of signal events expected at the Tera-Z run, while the latter will determine the search strategies to be adopted. As the design for the detector is not finalized, we will use the typical size of a central tracker to define the following three classes, based on the $a$ lifetime $L$:
\begin{itemize}
    \item[i)] For $L < 2$~cm, we consider the decays prompt.
    \item[ii)] For $2 < L < 100$~cm, we consider $a$ being long-lived as the decay off the beam pipe can be tracked and identified.
    \item[iii)] For $L>100$~cm, we consider the signature of $a$ to be missing energy, as it will decay outside the tracker and leave no identifiable trace in the detector.
\end{itemize}
The regions in the 2-dimensional parameter space corresponding to the above three classes are highlighted by the shaded regions in Fig.~\ref{fig:results} for both Photophobic and Photophilic ALPs.  Thanks to the prompt decay into photons, the Photophilic ALP has reduced regions with long-lived or missing energy phenomenology. For the Photophobic case, the prompt decays only occur above the $b\bar{b}$ threshold.

The Tera-Z phase of the FCC-ee and CepC colliders presents a unique opportunity for studying $Z$ portal physics. With the projected integrated Luminosity, $\sim 6\times 10^{12}$ $Z$ bosons will be produced, thus promising a remarkable reach for very rare $Z$ decays. In the models under consideration, we obtain branching ratios $Z\rightarrow a\gamma$ between $10^{-8}$ and $10^{-12}$, for values of  $f$ between $1$ and $100$~TeV . Thus, the Tera-Z run has the potential to probe compositeness scales $f$ well beyond what current electroweak precision and the LHC can do, as they probe $f$ up to a $2$--$3$ TeV~\cite{Cacciapaglia:2020kgq}.  Using a UFO model file from {\tt{FEYNRULES}} \cite{Alloul:2013bka}, we generated the matrix element of interest for the signal,  $\mathcal{M}(e^+e^-\rightarrow Z \rightarrow a\gamma)$, using {\tt{MADGRAPH}} \cite{Alwall:2011uj}. We pass the events through {\tt{PYTHIA}} for the showering and hadronization. The IDEA detector card of {\tt{DELPHES}} \cite{deFavereau:2013fsa} is used for the fast detector simulation. 

\begin{figure}[htb!]
	\centering	
	\begin{center}
		\begin{tabular}{c}
		\includegraphics[width=6.6cm,height=3.6cm]{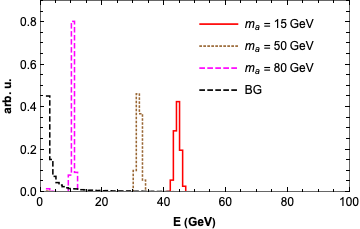} \\ \phantom{xxx} \\
		\includegraphics[width=6.6cm,height=3.65cm]{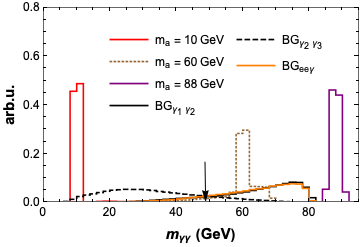}
		\end{tabular}
	\end{center}
	\caption{Illustration of the signal and background distribution for a given analysis.
	Top: energy of the leading isolated photon for the signal and the background. Bottom: construction of the invariant mass using di-photon events for different masses and the background. For $m_a\leq 50$ GeV, we use the two sub-leading photons ($\gamma^{(1)}\gamma^{(2)}$) while for other masses we use the two leading photons ($\gamma^{(0)}\gamma^{(1)}$). The corresponding background is illustrated by solid-black ($B_{\gamma_1\gamma_2}$) and dashed black ($B_{\gamma_2\gamma_3}$) line respectively.	}
	\protect\label{fig:distributions}
\end{figure}

\textit{Prompt decays.}
In the Photophobic models, prompt decays take place for masses above the $b\bar{b}$ threshold, where the dominant final state comprises bottom quarks. Thus we consider the main signal to be $Z \to \gamma a \to \gamma b \bar{b}$, leading to one isolated monochromatic photon and a pair of b-jets.
The jets are reconstructed using the AK4 \cite{Cacciari:2008gp} algorithm, requiring each jet to have a minimum transverse momentum $p_T^{min}=4$~GeV. Furthermore, we demand that at least one of the jets is tagged as a $b$-jet. Given the absence of data for a $p_T$ specific b-tagging efficiency for the IDEA detector, we assume a $p_T$-independent tag rate of $80\%$, with a mis-tag rate of $1\%$. The dominant background is due to the decay of $Z\rightarrow b\bar b \gamma$ where the photon is radiated off one of the b quarks (while $Z \to j j \gamma$ is suppressed by the jet mis-tag rate).
The energy of the monochromatic photon depends on the mass of the associated pseudo-scalar $a$, and can be used as the main discriminant against the background. 
In the top plot of Fig.~\ref{fig:distributions} we illustrate the distribution of the photon energy $E$ for three different masses of the pseudo-scalar $a$: $15$, $50$ and $80$~GeV.
The corresponding background distribution is shown by the black dashed line. We see that, as the mass of the pseudo-scalar increases, the energy of the photon moves closer to that of the background distribution. Hence, the differentiation of the signal will become increasingly ineffective. 
To quantify the discrimination power, and define our search strategy, we bin the events by $2$ GeV. We then use the following expression to evaluate the signal sensitivity \cite{Cowan:2010js}:
\begin{equation}
Z=\sqrt{\sum\limits_{i=1}^{N} \left(2(s_i+b_i)\log\left[1+\frac{s_i}{b_i}\right]-2s_i\right)}\,,
\label{eq:sensitivity}
\end{equation}
where the sum runs over all the bins and $s_i/b_i$ are signal/background events in the $i^{th}$ bin.
The right plot of Fig.~\ref{fig:results} shows the reach in the $(m_a,f)$ plane for the Photophobic composite models: in the blue-shaded region, the red and black contours show the $Z=2\sigma$ sensitivity for integrated Luminosities of $3$~ab$^{-1}$ and $150$~ab$^{-1}$, respectively. The contours clearly highlight the loss of sensitivity for masses close to the $Z$ mass, which is also due to a reduction in the B.R. of the $Z$ boson. With this channel we, therefore, expect to be sensitive to Higgs composite scales $f$ up to $5$~TeV, well above the current reach.

In the Photophilic models, the prompt region extends to masses as low as $1$~GeV, and the dominant decay produces two photons. The signal final state is $Z \to \gamma a \to \gamma \gamma \gamma$, thus containing three isolated photons. The photons are reconstructed with a minimum transverse momentum of $p_T^\gamma > 2$~GeV. Furthermore, we will order the photons in decreasing order of $p_T$ as $\gamma_1,\gamma_2,\gamma_3$.
We observe that for $a$ masses larger than $50$~GeV, in most of the events it's the two leading photons that reconstruct the resonance, while for masses below $50$~GeV it's the two sub-leading photons. Hence, we define a discriminant variable as
\begin{equation}
m_{\gamma \gamma} = \left\{ \begin{array}{l}
\left(p_{\gamma_1} + p_{\gamma_2}\right)^2 \;\; \mbox{for}\;\; m_a < 50~\mbox{GeV} \\
\left(p_{\gamma_2} + p_{\gamma_3}\right)^2 \;\; \mbox{for}\;\; m_a > 50~\mbox{GeV} 
\end{array} \right.
\end{equation}
depending on the $a$ mass we are probing. 
This variable is illustrated in the bottom plot of Fig.~\ref{fig:distributions} for three representative masses, showing that our strategy does indeed well reconstruct the resonance.
For the background, we first consider the irreducible process $e^+ e^- \to \gamma \gamma \gamma$. As for the signal, we define the background distribution in $m_{\gamma \gamma}$ by grouping either the leading (BG$_{\gamma_1\gamma_2}$) or sub-leading (BG$_{\gamma_2\gamma_3}$) photons, as shown by the solid and dashed black lines. The former is used for the computation of the signal sensitivity for masses $m_a\leq 50$~GeV, while the latter is used for the heavier masses. The blue arrow represents the transition point. 
An interesting feature in the BG$_{\gamma_1\gamma_2}$ distribution is its sharp fall above $80$~GeV. This can be attributed to the fact that the third photon carries a small but non-negligible energy. With a precise knowledge of the centre-of-mass energy and the fact that the three photons must necessarily reconstruct to mass around the $Z$ pole, the invariant mass of the two leading photons must exhibit falling distributions as one approaches the Z pole.
We also considered the background from $e^+ e^- \to e^+ e^- \gamma$, where both electrons are mis-identified as photons. The distribution of this background off the $Z$ pole is similar to that of the 3-photon one, and it can be kept below $2.5$\% of the irreducible background as long as the mis-id rate is kept below $0.0005$. To understand how reasonable this value is, we can compare with the study of a light-mass resonance decaying into photons in CMS \cite{Sirunyan:2018aui}, where the double-fake rate at the Z-pole was estimated to be $\sim 0.0015$. A more detailed knowledge of the detector and a data-driven analysis is required for a better understanding of this background. Similarly, the region for $m_{\gamma \gamma} > 80$~GeV will be populated dominantly by mis-id backgrounds, thus we remove this region from our analysis.
To quantify the sensitivity of this search, we use the same binned likelihood function in Eq.~\ref{eq:WZW} for $m_{\gamma \gamma}$. As before, the red and black contours correspond to $Z=2\sigma$ for $3$ and $150$~ab$^{-1}$. In the composite Higgs scale, this channel allows to probe $f$ up to $20$~TeV in the prompt decay region (blue shade). We checked that other bounds on the photon coupling \cite{Bauer:2017ris} only apply for $f<1$~TeV due to the loop suppression in Eq.~\eqref{eq:CW}.

\textit{Long-lived ALP signature.} 
Pseudo-scalars with a relatively long lifetime, such that a displaced vertex can be reconstructed, possibly give the most optimist picture due to the absence, in practice, of irreducible backgrounds. Yet, it is difficult to give reliable estimates without a precise definition of the detector. Here, therefore, we will provide an estimate of the number of events expected in each case assuming a negligible background. In the Photophobic case, the dominant decays involve hadrons, as shown in Fig.~\ref{fig:BF}. The contours in the orange region in the left plot of Fig.~\ref{fig:results} represent the benchmark of $2$ events with displaced hadrons $+ \gamma$ for an integrated luminosity of $3$ and $150$~ab$^{-1}$. 
This shows that the Tera-Z run has the potential of being sensitive to $f$ scales up to $600$~TeV.
For the Photophilic case, the long-lived signal can only occur for masses below $20$~GeV due to the leading coupling to photons. The signal thus consists of a monochromatic photon with energy between $35$ and $45$~GeV depending on the $a$ mass, and two photons originating from a displaced vertex. In this case too, we conservatively provide contours that give a sizeable number of events (20) for integrated luminosities of $3$ and $150$~ab$^{-1}$, as shown in the right plot of Fig.~\ref{fig:results}. The potential reach in $f$ is similar to the Photophobic case.

\textit{Missing energy signature.} 
For large life-times, which allow for decays outside of the tracker, the $a$ can be considered as missing energy as it will not be reconstructed in the detector. The signature, therefore, consists of a single monochromatic photon and has been studied in detail in Ref.~\cite{Cobal:2020hmk}. Here, decays of the $Z$ into a dark photon $\bar \gamma$ are considered, with a projected bound of  $\mathcal{B}(Z\rightarrow \gamma\bar\gamma)=2.3\times 10^{-11}$ for the Tera-Z.
Our case is slightly different, as the dark photon is replaced by a pseudo-scalar, however we do not expect large differences in the reach. Thus, we simply reinterpreted the bound in our parameter space, as shown by the horizontal red line in the green shaded regions of Fig.~\ref{fig:results}. The limit is the same for both models, as it is independent on the decay products of the $a$. 
It illustrates the sensitivity for scales as heavy as $\sim 90$ TeV.

\vspace{0.3cm}

In conclusion, we have considered the Tera-Z run at future $e^+ e^-$ colliders as a discovery machine for ALPs produced in the $Z$ decays. We consider composite Higgs models as a template, where the light ALP is a pseudo-Nambu-Goldstone boson that couples to the electroweak gauge bosons via the topological WZW term.
We consider both photophilic and photophobic scenarios, including loop-induced couplings to fermions and photons. Our analysis shows that the Tera-Z can probe the Higgs composite scale up to $5$ or $20$ TeV for prompt decays, and hundreds of TeV for long-lived or missing energy signatures.
Thus, the Tera-Z can zoom in the composite nature of the Higgs to much better precision than the LHC. 

Our results can be easily translated to other ALP scenarios. We remark that explanations of the muon $g-2$ anomaly \cite{Bauer:2017nlg}, recently confirmed at Fermilab \cite{Bennett:2006fi,Abi:2021gix}, need much larger couplings than the ones we consider here, therefore they can be severely constrained by searches in this channel.

\section*{Acknowledgements}

We are grateful to  Suzanne Gascon and Antoine Lesauvage for discussions on  ``double-fake'' events. We would like to thank Benjamin Fuks for his advice on b-tagging strategies. We acknowledge support from the CEFIPRA under the  project ``Composite Models at the Interface of Theory and Phenomenology'' (Project No. 5904-C).

\bibliography{biblio1}
\end{document}